\documentclass[reprint,prl,twocolumn,superscriptaddress,showpacs,nofootinbib,floatfix,preprintnumbers]{revtex4-1}
\usepackage{bm, 
  graphicx,
  hyperref,
  color,
  braket,
  array,
  slashed,
  mathrsfs,
  subfigure,
epsfig,amsmath,tkz-tab,multirow
}

\newcommand{\be}{\begin{eqnarray}}
\newcommand{\ee}{\end{eqnarray}}
\newcommand{\ba}{\begin{align}}
\newcommand{\ea}{\end{align}}

\begin{document}
\begin{flushright}
\end{flushright}

\title{Constraining Right Handed Gauge Boson Mass from Lepton Number Violating Meson Decays in a Low Scale Left Right Model}

\author{Sanjoy Mandal}
\email{smandal@imsc.res.in}
\affiliation{The Institute of Mathematical Sciences,
C.I.T Campus, Taramani, Chennai 600 113, India}
\affiliation{Homi Bhabha National Institute, BARC Training School Complex,
Anushakti Nagar, Mumbai 400094, India}
\author{Manimala Mitra\footnote{manimala@iopb.res.in}}
\email{manimala@iopb.res.in}
\affiliation{Institute of Physics, Sachivalaya Marg, Bhubaneswar 751005, India}
\author{Nita Sinha}
\email{nita@imsc.res.in}
\affiliation{The Institute of Mathematical Sciences,
C.I.T Campus, Taramani, Chennai 600 113, India}
\affiliation{Homi Bhabha National Institute, BARC Training School Complex,
Anushakti Nagar, Mumbai 400094, India}

\begin{abstract}
We analyze the lepton number violating (LNV) meson decays that arise in a TeV scale Left Right Symmetry model. The 
right handed Majorana neutrino $N$ along with the right handed or Standard Model gauge bosons  
mediate the meson decays and provide a resonant enhancement of the rates if the mass of $N$ ($M_N$) lies in the range $\sim (100\, \rm{MeV}-5\, \rm{GeV})$.
Using the expected upper limits on the number of events for the LNV decay modes $M_{1}^{+} \to\ell^{+}\ell^{+}\pi^{-}$~($M_{1}=B, D, D_{s}, K$), we derive constraints plausible on
the mass of the right handed charged gauge boson by future searches at the ongoing NA62 and LHCb experiments at CERN, the upcoming Belle II at SuperKEK, as well
as at the proposed future experiments, SHiP and FCC-ee.
These bounds are complimentary to the limits from same-sign dilepton search at Large Hadron Collider (LHC). The very high intensity of Charmed mesons expected
to be produced at SHiP will result in a far more stringent bound, $M_{W_R}>18.4$~TeV (corresponding to $M_N=1.46$~GeV), than the other existing bounds from  collider and
neutrinoless double beta decay searches.

\end{abstract}

\maketitle
\section{Introduction} \label{sec:I}
The observation of light neutrino masses and mixings provide unambiguous  experimental evidence for 
the existence of beyond standard model (BSM)  physics. So far, the solar and atmospheric  
mass square differences $\Delta m^2_{12}$, $|\Delta m^2_{13}|$ and the mixing angles $\theta_{12}$, $\theta_{23}$ and $\theta_{13}$ have been 
 measured with reasonable accuracy \cite{Gonzalez-Garcia:2015qrr}. {On the other hand, the cosmological constraints on the sum of light neutrino
 masses \cite{Ade:2015xua} guarantees the SM neutrino masses to be less than $\mathcal{O}(\rm{eV})$.} One of  the most attractive framework to explain the
 small light neutrino masses is the Minimal Left-Right Symmetry Model (MLRSM)\cite{LR}. The model offers several novel features including high scale parity
 symmetry, Majorana mass of the light and heavy neutrinos, explanation of  parity violation in  SM,  existence of the right handed current etc. The  light
 neutrino masses in this model are generated  from  dimension-five {lepton number violating (LNV) operator}~\cite{Weinberg1979sa} that violates lepton number
 by two units and hence their Majorana nature can be confirmed by observing the distinctive LNV signal at experiments, such as neutrinoless double beta
 decay ($0 \nu \beta \beta$) \cite{0nu2beta-old, schvalle, kamland, gerda}.  Additionally, the LNV signature can also be  tested at colliders from direct searches 
 \cite{KS, Chen:2013fna, Mitra:2016kov,Khachatryan:2014dka, Aad:2015xaa}, as well as through indirect searches from meson
 and tau decays~\cite{Atre:2009rg, Cvetic:2010rw, Aaij:2012zr, Mandal:2016hpr, Shuve:2016muy, Milanes:2016rzr}.
 While the light neutrinos can give dominant contribution in $0 \nu \beta \beta$, the LNV searches at collider and meson decays are however not sensitive
 to such small eV  mass scale. Hence, a positive result in the latter experiments will non-arguably prove the  existence of lepton number violating BSM states.

Contributions of BSM states of TeV or lighter masses to $0 \nu \beta \beta$, can be significantly large \cite{MSV}. The same feature is also applicable for a
TeV scale MLRSM, where the right handed neutrino $N$ together with a right handed gauge boson $W_R$ can give promising signals
in $0\nu \beta \beta$ and collider searches  \cite{Tello:2010,Barry:2013xxa,Dev:2013vxa, Dev:2014xea, Dell'Oro:2015tia, Ge:2015yqa, Awasthi:2016kbk, Meroni:2012qf}.
In addition, $W_R$ can also be looked for, through  dijet searches \cite{ATLAS:2015nsi, Khachatryan:2015dcf}. Right handed Majorana neutrinos with mass in the hundreds of
MeV-few GeV range, can be produced as an intermediate on mass shell state, resulting in a resonance enhancement of the LNV meson decay rates.
The detailed study of these  is the main objective of this paper. We follow a most generic
approach, taking into account all the contributions arising from right handed, left handed  currents,  as well as their combinations. We obtain constraints on the mass of $W_R$ that may be feasible from a number of ongoing and future experimental searches
	of meson decay modes with like sign dileptons, such as $K\to\ell\ell\pi$, $D_s\to\ell\ell\pi$, $D\to\ell\ell\pi$ and $B\to\ell\ell\pi$.
	The huge number of $D_s$ meson decays expected in the SHiP experiment will result in the most stringent constraint on $W_R$ mass, corresponding to $M_N=1.46$~GeV.

The paper is organized as follows: we first review the basic features of the MLRSM, following which we discuss in detail the contributions of the right
handed neutrino and right/left handed gauge bosons in meson decays. The total decay width of the heavy Majorana neutrino having a mass in between that of the
pion and the $B$ meson, is computed. We then derive the limits on the mass of $W_R$ that are expected from the upper limits on the number of events of various LNV meson decays that may be achievable in
some of the ongoing and future experiments. Finally we provide our conclusions.
In the Appendix, we provide some details of the calculations.

\section{Left-Right Symmetric Model \label{model}}
The minimal Left Right Symmetric Model is based on the  gauge group $SU(3)_c\times SU(2)_L\times SU(2)_R\times U(1)_{B-L} $\cite{LR},
with the fermions assigned in the doublet representation of $SU(2)_L$ and $SU(2)_R$. In addition to the particle content of the Standard Model (SM),
the model contains three right handed Majorana neutrinos {$N_R$}, and the   additional gauge bosons $W_R$ and $Z^{\prime}$. The electric charge $Q$ and the third
components of  weak isospins $I_{3L}$ and $I_{3R}$  are related as  $Q=I_{3L}+I_{3R}+(B-L)/2$. The scalar sector of this model consists of the bi-doublet
$\Phi$ and the Higgs triplets $\Delta_L$ and $\Delta_R$, where the Higgs states have the following representations:
$\Phi ({\bf 1}, {\bf 2}, {\bf 2}, 0)$,  $\Delta_L (1, 3,1,+2)$ and $\Delta_R (1, 1,3,+2)$. The  bi-doublet, being neutral under $B-L$ gauge group is not sufficient to break {this gauge symmetry}. Hence, additional Higgs triplet fields are required. {{The Higgs field $\Delta_R$ takes vacuum expectation value $v_R$ and breaks $SU(2)_R\times U(1)_{B-L}$  down to the group $U(1)_Y$ of SM. }}

{In the Yukawa sector,} the bi-doublet couples  to the fermion bilinears $\bar{Q}_LQ_R$ and $\bar{\psi}_L\psi_R$, and gives masses to quarks and leptons {through the }
spontaneous symmetry breaking, {{where its VEVs are denoted as}}: $\langle\Phi\rangle={\rm diag}(\kappa_1, \kappa_2)/\sqrt{2}$. {On the other hand, the Higgs triplet
$\Delta_R$ couples with the right handed neutrinos $N_R$ and  generates the Majorana mass of heavy neutrinos during the symmetry breaking. While the heavy neutrino $N_R$
contributes to the light neutrino mass generation via  Type-I seesaw mass \cite{type1,type1b}, the triplet Higgs  $\Delta_L$ generates the Majorana mass of light neutrinos
via Type-II seesaw \cite{type2}. The VEV of $\Phi$ field breaks} the SM gauge group $SU(2)_L\times U(1)_Y$ to $U(1)_Q$. Hence, the different {VEVs of bi-doublets and
triplets follow} the hierarchy $v_L\ll \kappa_{1,2} \ll v_R \;
\label{eq:hie}$.

\begin{figure}[]
\subfigure[]{\includegraphics[width=8cm]{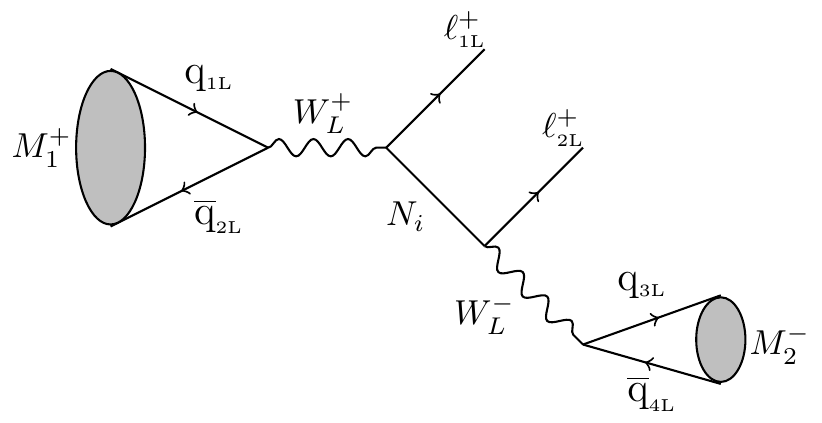}}
\subfigure[]{\includegraphics[width=8cm]{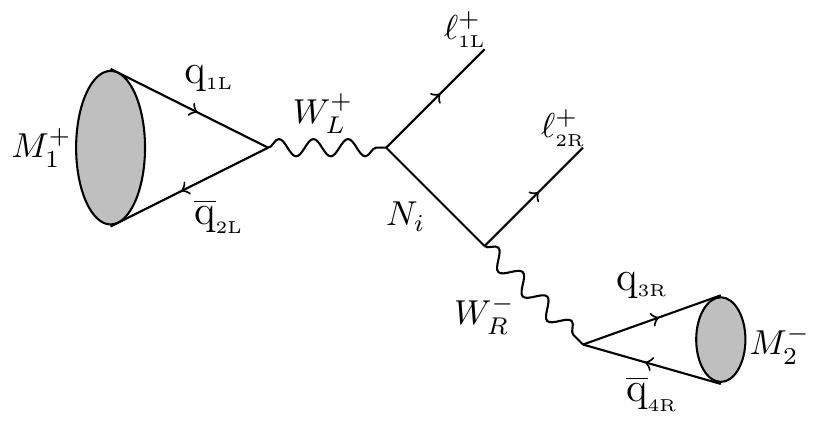}}
\subfigure[]{\includegraphics[width=8cm]{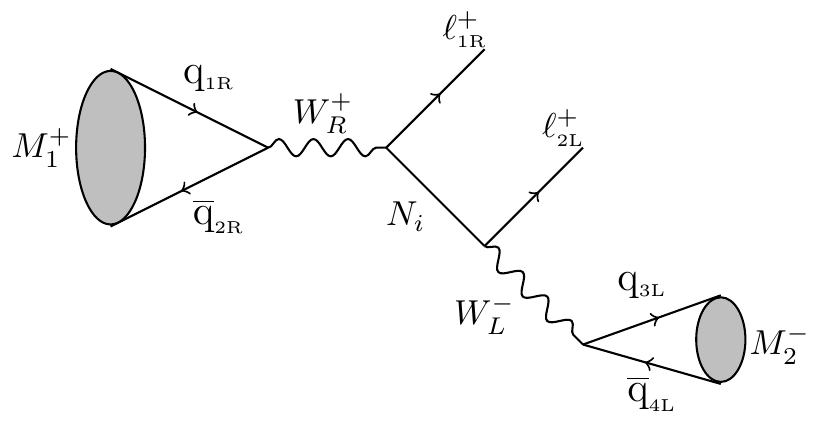}}
\subfigure[]{\includegraphics[width=8cm]{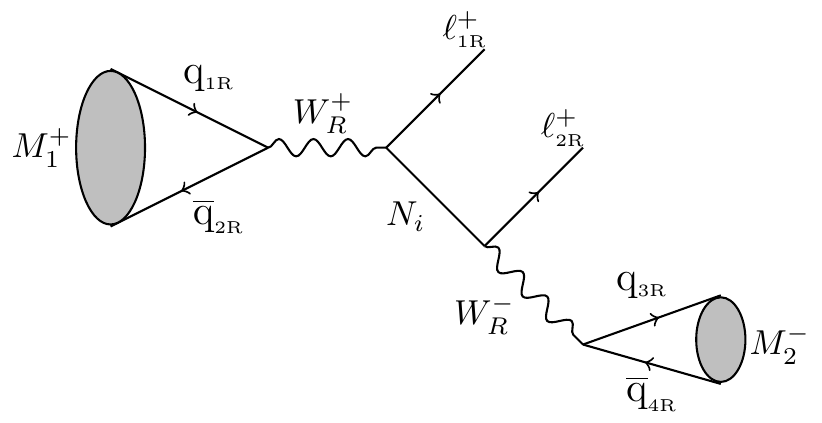}}
\caption{\small{The Feynman diagrams for the lepton number violating 
meson decays. These processes produce resonance enhancement. See text for details.}}
\label{Feynman diagrams1}
\end{figure}

\begin{figure}[]
\subfigure[]{\includegraphics[width=8cm]{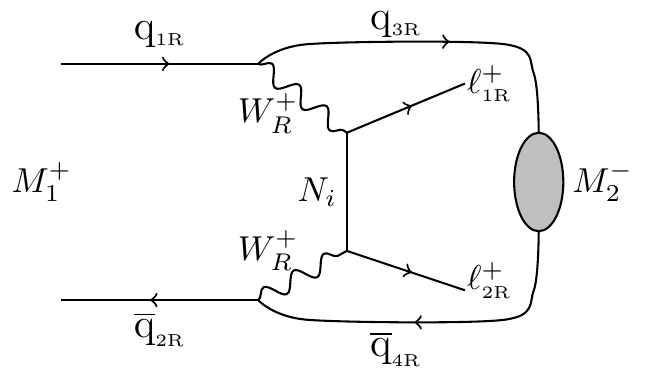}}
\caption{\small{The t-channel diagram for the lepton number violating meson decay. See text for details.} }
\label{Feynman diagrams2}
\end{figure}

The Yukawa Lagrangian, that generates the lepton masses have the following form:
\begin{eqnarray}
-{\cal L}_Y &=& h\bar{\psi}_{L}\Phi \psi_{R} 
+ \tilde{h} \bar{\psi}_{L}\tilde{\Phi} \psi_{R}
+ f_{L} \psi_{L}^{\sf T} C i\tau_2 \Delta_L \psi_{L} \nonumber \\ 
&& + f_{R} \psi_{R}^{\sf T} C i\tau_2 \Delta_R \psi_{R} 
+{\rm H.c.}
\label{eq:yuk}
\end{eqnarray}
In the above $C$ is the charge-conjugation matrix,  $C=i\gamma_2\gamma_0$, and $\tilde{\Phi}=\tau_2\Phi^*\tau_2$, with $\tau_2$ being the second Pauli matrix,
and $\gamma_\mu$ the Dirac matrices. Upon symmetry breaking, this gives rise to the following light-heavy mass matrix,
\begin{eqnarray}
{\cal M}_\nu = \left(\begin{array}{ccc}
M_L & M_D \\
M_D^{\sf T} & M_R
\end{array}\right).
\label{eq:big}
\end{eqnarray} In the seesaw approximation \cite{type1,type1b,type2}, this leads to the following light neutrino mass matrix  (up to $\mathcal{O}(M^{-2}_R)$) \cite{Grimus:2000vj},
\begin{eqnarray}
M_\nu  & \simeq &  M_L - M_D M_R^{-1} M_D^{\sf T}- \nonumber \\
&& (\frac{1}{2} M_D M^{-1}_R {M^{-1}_R}^* M^{\dagger}_D M_L+\rm{HC}) \nonumber \\
 &&  =  \sqrt 2 v_L f_L - \frac{\kappa^2}{\sqrt 2 v_R} h_D f_R^{-1} h_D^{\sf T}- \nonumber \\
&&( \frac{\kappa^2}{2 \sqrt 2 v^2_R v_L} h_D f_R^{-1}{f_R^{-1}}^* h_D^{\sf \dagger}f_L+h.c)\, \;,
\label{eq:mnu}
\end{eqnarray}
where the Dirac mass $M_D = h_D \kappa=\frac{1}{\sqrt 2}\left(\kappa_1 h + \kappa_2 \tilde{h} \right)$, 
$\kappa= \sqrt{\kappa^2_1+\kappa^2_2}$ and $
M_L = \sqrt 2 v_L f_L$, $M_R = \sqrt 2 v_R f_R $. The mass matrix given in Eq.~(\ref{eq:big}), can be diagonalized 
 by a $6\times 6$ unitary matrix, as follows: 
\be
{\cal V}^{\sf T}{\cal M}_\nu {\cal V} \ = \ \left(\begin{array}{cc} \widetilde{M}_\nu & {\bf 0} \\ {\bf 0} & \widetilde{M}_R \end{array}\right) \ee
where ${\widetilde{M}_\nu}={\rm{diag}}(m_1,m_2,m_3)$,  $\widetilde{M}_R = {\rm diag}(M_1,M_2,M_3)$. Upto $\mathcal{O}(M^{-2}_R)$, the mixing matrix $\cal{V}$ has
the following form, 
\be
\cal{V} & \sim & \left(\begin{array}{cc}
{\bf 1-\frac{1}{2}\zeta^* \zeta^T} & \zeta^*+{\zeta^{\prime}}^* \\
-\zeta^{\sf T}-{\zeta^{\prime}}^{\sf T} & {\bf 1-\frac{1}{2} \zeta^T \zeta^{*}}
\end{array}  \right)
\left(\begin{array}{cc}
U_\nu & {\bf 0} \\
{\bf 0} & V_R
\end{array}\right).
\label{V}
\ee
In the above,  the expansion parameter $\zeta$ has the following form $\zeta=M_DM_R^{-1}$, and ${\zeta^{\prime}}^*=M^{\dagger}_LM_DM^{-1}_R{M^{-1}_R}^*$.
For $M_L\rightarrow 0$, $\zeta^{\prime}\rightarrow 0$. The order parameter is defined as, $\theta\equiv ||\zeta||=\sqrt{Tr(\zeta^\dagger\zeta)}$. In the subsequent
analysis, we denote the mixing matrix as 
\be
\cal{V} &= &
\left(\begin{array}{cc}
 U & S \\
T & V 
\end{array} \right).
\label{mat}
\ee

 \subsection{ Gauge Sector and Charged Current Lagrangian}

In addition to the SM gauge bosons, this model consists of a right handed charged gauge boson, $W_R$ and an additional neutral gauge boson, $Z'$. The SM gauge boson 
$W_L$ and $W_R$ mix with each other, with the mixing \be
\xi \simeq \frac{\kappa_1\kappa_2}{v_R^2} \simeq \frac{2\kappa_2}{\kappa_1}\left(\frac{M_{W_L}}{M_{W_R}}\right)^2. 
\label{xiup}
\ee
In the limit of small mixing $\xi \ll 1$,  the  physical masses are 
\begin{eqnarray}
M_{W_1} \simeq M_{W_L} \simeq \frac{g}{2}\kappa, \, \,\,  M_{W_2} \simeq M_{W_R} \simeq \frac{g}{\sqrt 2}v_R~,
\end{eqnarray}
where, $g\equiv g_L=g_R$ and $M_{Z^\prime} \sim 1.7 M_{W_R}$. In our choice of Left-Right model, we assume discrete parity as a symmetry. 
The charged current Lagrangian for the quarks  have the following forms: 
\begin{eqnarray}
\mathcal{L}^q_{CC} &=& \frac{g}{\sqrt{2}}{\sum_{i,j}}\overline{u}_i V_{ij}^{\rm CKM}~W_{L \mu}^+ \gamma^\mu P_L~ d_j+ \nonumber \\ 
&& \frac{{g}}{\sqrt{2}}\sum_{i,j}\overline{u}_i V_{ij}^{\rm R-CKM'}~W_{R \mu}^+ \gamma^\mu P_R~ d_j +   \text{H.c.},
\label{cc1}
\end{eqnarray}
where $i= (u,c, t)$ correspond to the up type quarks and {$j=(d,s,b)$} represent the down type quarks. In our subsequent analysis, we consider $V^{\rm{R-CKM'}}$
to be proportional to $V^{\rm{CKM}}$ with the proportionality factor  $\beta \sim \mathcal{O}(1)$. The charged current Lagrangian for the  lepton-neutrino has the following form, 
\begin{eqnarray}
 \mathcal{L}^\ell_{CC} &=& \frac{g}{\sqrt{2}} \sum_{i,j}\overline{\ell}_{L_i}  W_{L \mu}^- \gamma^\mu P_L~ ( U_{ij} \nu_{L_j} + S_{ij} N^c_{j})+ \nonumber \\ && \frac{{g}}{\sqrt{2}}\sum_{i,j}\overline{\ell}_{R_i} ~W_{R \mu}^- \gamma^\mu P_R~ (V^*_{ij}{N_{j}+T^*_{ij} \nu_L^c)} +  \nonumber \\ &&  \text{H.c.}
\label{cc2}
\end{eqnarray}

Note that, both the masses of the RH neutrino and $W_R$ gauge boson, are proportional to the $SU(2)_R$ breaking scale. However, while the $W_R$ mass
and the {right handed current} are dictated with the gauge coupling $g_R=g$, the RH neutrino mass $M_N$ is governed by the Yukawa coupling {$f_R$,
therefore allowing the possibility to have a large hierarchy between $W_R$ and $N$ masses}. For sufficiently small $M_N$ and TeV scale $W_R$, this
can give large contribution in meson decays. In this analysis, we do not restrict ourselves to a particular parameter range. Instead we consider
the possibility that the heavy neutrino  $N$ is of 100 MeV-5 GeV scale, {where} the meson decays can be resonantly enhanced. { We follow a generic
approach for the calculation of branching ratios by taking into account all the contributions that can originate from MLRSM}. 

We show  the complete list of decay modes of $N$ along with the corresponding rates in the Appendix. The neutral currents which will also contribute in some of the decay channels of
the right handed neutrino, are given below:~\cite{Tello:2012qda, Mohapatra:1977be} 
\begin{eqnarray}
\mathcal{L}_{NC}=\frac{g_L}{cos \theta_w}(Z_{\mu}J^{\mu}_{Z}+\frac{cos^2 \theta_w}{\sqrt{cos 2\theta_w}}Z_{\mu}^\prime J^{\mu}_{Z^\prime})
\label{nc}
\end{eqnarray}
where, 
\begin{eqnarray}
J^{\mu}_{Z}=\sum_i\bar{f}\gamma^{\mu}(T^3_LP_L-Q sin^2 \theta_w)f, \\
J^{\mu}_{Z^\prime}=\sum_i\bar{f}\gamma^{\mu}(T^3_RP_R-tan^2 \theta_w(Q-T^3_{3L}) f.
\label{currents}
\end{eqnarray}

The explicit interaction terms with the leptons and neutrinos are given in \cite{Tello:2012qda}.

\section{Imprint of Majorana Signature in Meson Decays   \label{mes1}}
In the simplest LRSM, the heavy neutrinos $N_{i}$ are Majorana, that 
inherently carry lepton number violation. Together with the gauge bosons $W_R$, or even with $W_L$, they can
mediate the lepton number violating meson decays, $M_{1}^{+}(p)\rightarrow\ell_{1}^{+}(k_{1})\ell_{2}^{+}(k_{2})M_{2}^{-}(k_{3})$, where $M_1$ is a pseudoscalar,
while $M_2$ can be a pseudoscalar or a vector meson. We assume that there are three RH neutrinos
with masses in the $100 \, \text{MeV}-5$ GeV range, 
that contribute in these meson decays. The Feynman diagrams for these decays are shown in Figs~\ref{Feynman diagrams1} and ~\ref{Feynman diagrams2}.
The different contributions are
mediated through $W_L-N_{i}-W_L$ (Fig.~\ref{Feynman diagrams1}(a)), while those in Fig.~\ref{Feynman diagrams1}(b) and Fig.~\ref{Feynman diagrams1}(c) are mediated by $W_L-N_{i}-W_R$ and
$W_R-N_{i}-W_L$, respectively. All these contributions depend on the active-sterile neutrino mixing {$S_{\ell_{j}N_{i}}$} and the RH neutrino mixing $V_{\ell_{j}N_{i}}$, while the diagram shown in Fig ~\ref{Feynman diagrams1}(d),
is mediated with $W_R-N_{i}-W_R$, and depends on $V_{\ell_{j}N_{i}}$.

The diagram in Fig.~\ref{Feynman diagrams2} (and a similar diagram with two $W_L$'s, as well as the diagrams with doubly charged Higgs triplets exchange) will give a
small contribution, as this is not a s-channel resonance production diagram. In addition, there can also be additional diagrams with $W_L-W_R$ mixing in one
of the legs. These however, will be small compared to the diagrams discussed above, as these come with  a further suppression factor of $\tan \xi$, due to the
$W_L-W_R$ mixing.  Hence, we do not consider these in our analysis. Further, note that, the contributions from light neutrino exchange will be negligibly small as
they will not be resonantly enhanced.
 
Below, we explicitly write the
amplitudes for the LNV decays of pseudoscalar mesons to a final pseudoscalar as well as to a vector meson. 
For each of these decays, the $LL$, $RR$, $LR$ and $RL$ contributions are a sum of two terms, where the second term is obtained by interchanging the
momenta $k_1$ with $k_2$ of the 2 leptons, as well as interchanging the leptonic mixing elements. Hence for decay to a pseudoscalar (vector) meson we may write,

\begin{eqnarray}
 \mathcal{M}_{h_1 h_2}^{P(V)}=\mathcal{M}_{1h_1 h_2}^{P(V)}+\mathcal{M}_{2h_1 h_2}^{P(V)} \nonumber 
 \end{eqnarray}
where, $h_1 h_2$ can be of {different chiralities} $LL, RR,LR,RL$.

\begin{widetext}
 \begin{eqnarray} 
\mathcal{M}_{1LL}^{P}=\sum_{i}G_{F}^{2}V_{M_{1}}^{CKM}V_{M_{2}}^{CKM} f_{M_{1}} f_{M_{2}}M_{N_{i}}\left(S_{\ell_{1}N_{i}}^{*}S_{\ell_{2}N_{i}}^{*}\right)
 \frac{\overline{u}(k_{2})\slashed{k_{3}}\slashed{p}\left(1-\gamma_{5}\right)v(k_{1})}{\left(p-k_{1}\right)^{2}-M_{N_{i}}^{2}+iM_{N_{i}}\Gamma_{N_{i}}}
\label{ll}
\end{eqnarray}
\begin{eqnarray}
 \mathcal{M}_{1RR}^{P}=\sum_{i}G_{F}^{2}V_{M_{1}}^{CKM}V_{M_{2}}^{CKM} f_{M_{1}} f_{M_{2}}M_{N_{i}}\left(\frac{M_{W_{L}}^{4}}{M_{W_{R}}^{4}}
 \right)(V_{\ell_{1}N_{i}}V_{\ell_{2}N_{i}})
 \frac{\overline{u}(k_{2})\slashed{k_{3}}\slashed{p}\left(1+\gamma_{5}\right)v(k_{1})}{\left(p-k_{1}\right)^{2}-M_{N_{i}}^{2}+iM_{N_{i}}\Gamma_{N_{i}}}~.
 \label{rr}
\end{eqnarray}
\end{widetext}
{where $k_3$ and $p$ are the four momentums of $M^+_2$ and $M^-_1$ mesons}. The $LR$ and $RL$ contributions are 
\begin{widetext}
\begin{eqnarray}
 \mathcal{M}_{1LR}^{P}=\sum_{i}G_{F}^{2}V_{M_{1}}^{CKM}V_{M_{2}}^{CKM} f_{M_{1}} f_{M_{2}}\left(\frac{M_{W_{L}}^{2}}{M_{W_{R}}^{2}}\right)(S_{\ell_{1}N_{i}}^{*}V_{\ell_{2}N_{i}})
 \frac{\overline{u}(k_{2})\slashed{k_{3}}\left(\slashed{p}-\slashed{k}_{1}\right)\slashed{p}\left(1-\gamma_{5}\right)v(k_{1})}{\left(p-k_{1}\right)^{2}-M_{N_{i}}^{2}+iM_{N_{i}}\Gamma_{N_{i}}}
\label{eqlr}
\end{eqnarray}
\begin{eqnarray}
  \mathcal{M}_{1RL}^{P}=\sum_{i}G_{F}^{2}V_{M_{1}}^{CKM}V_{M_{2}}^{CKM} f_{M_{1}} f_{M_{2}}\left(\frac{M_{W_{L}}^{2}}{M_{W_{R}}^{2}}\right)(V_{\ell_{1}N_{i}}S_{\ell_{2}N_{i}}^{*})
 \frac{\overline{u}(k_{2})\slashed{k_{3}}\left(\slashed{p}-\slashed{k}_{1}\right)\slashed{p}\left(1+\gamma_{5}\right)v(k_{1})}{\left(p-k_{1}\right)^{2}-M_{N_{i}}^{2}+iM_{N_{i}}\Gamma_{N_{i}}}~.
\label{eqrl}
\end{eqnarray}
In the above,  the decay rate $\Gamma^{P}$ is 
\end{widetext}
\begin{widetext}
\begin{equation}
\Gamma^{P}\left(M_{1}\rightarrow\ell_{1}\ell_{2}M_{2}\right)=\frac{1}{n!}\left(\left|\mathcal{M}_{LL}^{P}+\mathcal{M}_{RR}^{P}+\mathcal{M}_{LR}^{P}+\mathcal{M}_{RL}^{P}\right|^{2}\right)d_{3}(PS)~,
\label{eqgamma}
\end{equation}
\end{widetext}

In Eqs.(\ref{ll})-~(\ref{eqrl}),  $G_{F}$ is the Fermi coupling constant, {$V_{\ell_{j}N_{i}}$  are the elements of the mixing matrix for $N_{i}$,
$S_{\ell_{j}N_{i}}$ are the }elements between the neutrino of flavor state $\nu_{\ell_{j}}$ and mass eigenstate $N_{i}$, 
$V_{M_{1}}^{CKM}$, $V_{M_{2}}^{CKM}$ are the Cabbibo-Kobayashi-Maskawa (CKM) matrix elements at the annihilation
(creation) vertex of the meson $M_1$($M_2$),  $f_{M_{1}}$, $f_{M_{2}}$ are the decay constants of $M_{1}$, $M_{2}$, and $M_{N_{i}}$, $\Gamma_{N_{i}}$ are
the mass and decay width of the heavy 
neutrino $N_{i}$.
For the case of a vector meson in the final state, the amplitudes are as follows:

\begin{widetext}
\begin{eqnarray}
\mathcal{M}_{1LL}^{V}=\sum_{i}G_{F}^{2}V_{M_{1}}^{CKM}V_{M_{2}}^{CKM} f_{M_{1}} f_{M_{2}}M_{N_{i}}m_{3}\left(S_{\ell_{1}N_{i}}^{*}S_{\ell_{2}N_{i}}^{*}\right)\nonumber
 \frac{\overline{u}(k_{2})\slashed{\epsilon}^{\lambda}(k_{3})\slashed{p}\left(1-\gamma_{5}\right)v(k_{1})}{\left(p-k_{1}\right)^{2}-M_{N_{i}}^{2}+iM_{N_{i}}\Gamma_{N_{i}}}\end{eqnarray}
\begin{eqnarray}
\mathcal{M}_{1RR}^{V}=\sum_{i}G_{F}^{2}V_{M_{1}}^{CKM}V_{M_{2}}^{CKM} f_{M_{1}} f_{M_{2}}M_{N_{i}}m_{3}\left(\frac{M_{W_{L}}^{4}}{M_{W_{R}}^{4}}\right)(V_{\ell_{1}N_{i}}V_{\ell_{2}N_{i}})
 \frac{\overline{u}(k_{2})\slashed{\epsilon}^{\lambda}(k_{3})\slashed{p}\left(1+\gamma_{5}\right)v(k_{1})}{\left(p-k_{1}\right)^{2}-M_{N_{i}}^{2}+iM_{N_{i}}\Gamma_{N_{i}}}\end{eqnarray}
\end{widetext}
The $LR$ contributions are 
\begin{widetext}
\begin{eqnarray}
 \mathcal{M}_{1LR}^{V}=\sum_{i}G_{F}^{2}V_{M_{1}}^{CKM}V_{M_{2}}^{CKM} f_{M_{1}} f_{M_{2}}m_{3}\left(\frac{M_{W_{L}}^{2}}{M_{W_{R}}^{2}}\right)(S_{\ell_{1}N_{i}}^{*}V_{\ell_{2}N_{i}})
 \frac{\overline{u}(k_{2})\slashed{\epsilon}^{\lambda}(k_{3})\left(\slashed{p}-\slashed{k}_{1}\right)\slashed{p}\left(1-\gamma_{5}\right)v(k_{1})}{\left(p-k_{1}\right)^{2}-M_{N_{i}}^{2}+iM_{N_{i}}\Gamma_{N_{i}}}\nonumber\\
\end{eqnarray}
\begin{eqnarray}
 \mathcal{M}_{1RL}^{V}=\sum_{i}G_{F}^{2}V_{M_{1}}^{CKM}V_{M_{2}}^{CKM} f_{M_{1}} f_{M_{2}}m_{3}\left(\frac{M_{W_{L}}^{2}}{M_{W_{R}}^{2}}\right)(V_{\ell_{1}N_{i}}S_{\ell_{2}N_{i}}^{*})
 \frac{\overline{u}(k_{2})\slashed{\epsilon}^{\lambda}(k_{3})\left(\slashed{p}-\slashed{k}_{1}\right)\slashed{p}\left(1+\gamma_{5}\right)v(k_{1})}{\left(p-k_{1}\right)^{2}-M_{N_{i}}^{2}+iM_{N_{i}}\Gamma_{N_{i}}}\nonumber\\
\end{eqnarray}
\end{widetext}
In the above, $\epsilon^{\lambda}(k_{3})$ is the polarization vector of meson $M_{2}$ and $k_3^2=m_3^2$. The other contributions $\mathcal{M}_{2LL}^{V},
\mathcal{M}_{2RR}^{V},\mathcal{M}_{2LR}^{V}$ and $\mathcal{M}_{2RL}^{V}$ can be obtained by interchanging $k_1$ with $k_2$.

\begin{figure}
\hspace*{-0.7cm}
\includegraphics[width=10cm]{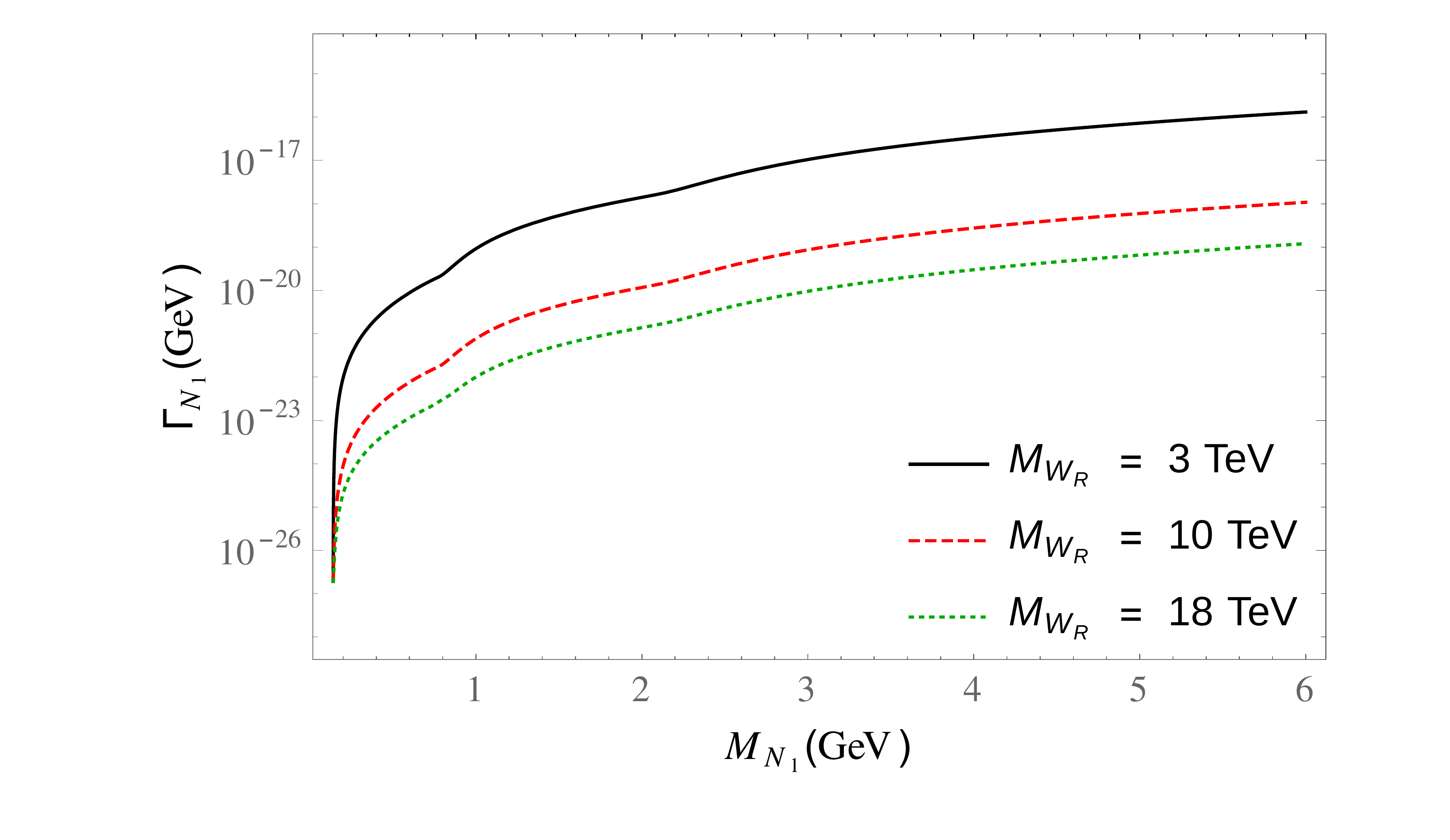}
\caption{The total decay width of the heavy neutrino $N_{1}$.}
\label{fig:figgamma}
\end{figure}

\section{Total Decay Width of the heavy Majorana neutrino N}
If the mass of $N_{i}$ lies in the range $0.140~\rm{GeV}<M_{N_{i}}<5.3~\rm{GeV}$, it can be produced as an intermediate on mass shell state in the lepton number
violating  meson decay modes being considered. { We compute the  total width of $N_{i}$ including all possible decay channels, that are  listed below.} We
consider only  tree level diagrams for {the computation.} The explicit expressions of the partial decay widths are given in the {Appendix.} In addition
to the SM gauge bosons $W_L$, $Z$, the gauge bosons $W_R$, $Z'$ will also contribute in the following two and three body decays of RH neutrinos via charged
current and neutral current interactions.

\begin{itemize}
\item
 RH neutrino decays to a charged pseudoscalar meson: $N_{i}$$\to\ell^{-}P^{+}$, where $\ell=e,\mu,\tau$ and $P^{+}$$=\pi^{+},\,K^{+},\,D^{+},\,D_{s}^{+}$. 
\item
RH neutrino decays to a neutral pseudoscalar meson:  $N_{i}\rightarrow\nu_{\ell}P^{0}$, where $\nu_{\ell}$ are the flavor eigenstates $\nu_{e},\,\nu_{\mu},\,\nu_{\tau}$ and $P^{0}=\pi^{0},\,\eta,\,\eta',\,\eta_{c}$.
\item
RH neutrino decays to a charged vector meson: $N_{i}\rightarrow\ell^{-}V^{+}$, where $\ell~=~e,\,\mu,\,\tau$ and $V^{+}=\rho^{+},
K^{*+},\,D^{*+},\,D_{s}^{*+}$.
\item
RH neutrino decays to neutral vector meson: $N_{i}\rightarrow\nu_{\ell}V^{0}$, where $\nu_{\ell}=\nu_{e},\,\nu_{\mu},\,\nu_{\tau}$ and $V^{0}=\rho^{0},\,\omega,\,\phi,\,J/\psi$. \\ 
\begin{center} Three body leptonic decays of $N_{i}$ \end{center} 

\item $N_{i}\rightarrow\ell_{1}^{-}\ell_{2}^{+}\nu_{\ell_{2}}$, where $\ell_{1},\,\ell_{2}=e,\,\mu,\,\tau$, $\ell_{1}\neq
\ell_{2}.$
\item $N_{i}\rightarrow\nu_{\ell_{1}}\ell_{2}^{-}\ell_{2}^{+}$, where $\ell_{1},\,\ell_{2}=e,\,\mu,\,\tau$. 
\item $N_{i}\rightarrow\nu_{\ell_{1}}\nu\overline{\nu}$, where $\nu_{\ell_{1}}=\nu_{e},\,\nu_{\mu},\,\nu_{\tau}$.
\end{itemize}

The total decay width of heavy majorana neutrino $N_{i}$ is given by
\begin{equation}
\begin{split}
&\Gamma_{N_{i}}=\sum_{\ell,P}2\Gamma^{\ell P}+\sum_{\ell,P}\Gamma^{\nu_{\ell}P}+\sum_{\ell,V}2\Gamma^{\ell V}+\sum_{\ell,V}\Gamma^{\nu_{\ell}V}\\
&+\sum_{\ell_{1},\ell_{2}(\ell_{1}\neq\ell_{2})}2\Gamma^{\ell_{1}\ell_{2}\nu_{\ell_{2}}}+\sum_{\ell_{1},\ell_{2}}\Gamma^{\nu_{\ell_{1}}\ell_{2}\ell_{2}}+\sum_{\nu_{\ell_{1}}}\Gamma^{\nu_{\ell_{1}}\nu\overline{\nu}}.
\end{split}
\end{equation}
	As each of $N_{i}$ are Majorana, they can also decay to the charge conjugate of the decay modes $N_{i}\rightarrow\ell^{-}P^{+}$,
	$N_{i}\rightarrow\ell^{-}V^{+}$, $N_{i}\rightarrow\ell_{1}^{-}\ell_{2}^{+}\nu_{\ell_{2}}$ with the same partial width,
	resulting in a 2 factor associated with these widths. In deriving the above relations, we neglect the contributions from the small mixing
	between $W_L-W_R$ and $Z-Z'$. We show the total decay width of the heavy neutrino $N$ in Fig.~\ref{fig:figgamma} for different choices of $W_R$ masses.
	In our analysis we consider the following mixing texture:
	
	\begin{itemize}
		\item
		$U_{\nu}=U_{\text{PMNS}}$.
		\item
		$V_R=I$, 
	\end{itemize}
	i.e., the RH neutrinos $N_{i}$'s are in the mass basis. The numerical values of the elements of the PMNS mixing matrices taken from Ref.~\cite{Gonzalez-Garcia}.

\section{Limit on $M_{W_R}$ from {{ongoing}} and future experiments in presence of heavy Majorana neutrinos} 

We consider the LNV signatures from decay modes $M_{1}^{+}\to\ell^{+}\ell^{+}M_{2}^{-}$. As stated before, the contributions from  LL, RL, LR diagrams depend on the active-sterile neutrino mixing $S$,
whereas the RR contribution depends on the RH mixing matrix $V_R$. 
The active-sterile mixing from Eq.~(\ref{V}) is {$S=\theta V_R$}. Without loss of generality, one can approximate $\theta \sim \sqrt{\frac{m_{\nu}}{M_N}}$, where $m_{\nu}$ and $M_N$ are
the light and RH neutrino masses, in accordance with the seesaw condition.

The sensitivity reach for the above LNV decay modes in a particular experiment depends on the number of the parent mesons $M_{1}$'s produced
($N_{M_1^+}$), their momentum ($\vec{p}_{M_{1}}$) 
 and the branching ratio for these mesons to the LNV modes. Assuming the parent meson $M_1$ decays at rest, 
the expected number of signal events is~\cite{Asaka}:
\begin{align}\label{event signal in rest}
N_{event}=2N_{M_{1}^{+}}\text{Br}\left(M_{1}^{+}\to\ell^{+}\ell^{+}M_{2}^{-}\right)\mathcal{P}_{N},\nonumber\\
\approx 2N_{M_{1}^{+}}\text{Br}\left(M_{1}^{+}\to\ell^{+}N_{i}\right)\frac{\Gamma(N_{i}\to\ell^{+}M_{2}^{-})}{\Gamma_{N_{i}}}\mathcal{P}_{N}, 
\end{align}
where  for $\ell=e$, $N_{i}=N_{1}$ and for $\ell=\mu$, $N_{i}$=$N_{2}$, the factor 2 is due to inclusion of the charge conjugate process
$M^{+}_1 \to \ell^{+} N_{i}$ and $\mathcal{P}_{N}$, is the probability of the RH neutrino $N_{i}$ to  decay within a detector of the length $L_{D}$ given by:
\begin{align*}
\mathcal{P}_{N}=\left[1-exp\bigg(-\frac{M_{N_{i}}\Gamma_{N_{i}}L_{D}}{p^*_{N_{i}}}\bigg)\right].
\end{align*}
In the above,  $p^*_{N_{i}}=\frac{m_{M_{1}}}{2}\lambda^{\frac{1}{2}}\big(1,\frac{m_{\ell}^{2}}{m_{M_{1}}^{2}},\frac{M_{N_{i}}^{2}}{m_{M_{1}}^{2}}\big)$ is
the momentum of $N_{i}$ in $M_{1}$ rest frame. For  the meson $M_{1}$ produced with fixed boost $\vec{\beta}$, the energy of $N_{i}$ is then given by,
\begin{align*}
E_{N_{i}}=E_{N_{i}}^{*}\left(\gamma+\frac{p_{N_{i}}^{*}}{E_{N_{i}}^{*}}\sqrt{\gamma^{2}-1}~\text{cos}~\theta_{N_{i}}^{*}\right),
\end{align*}
where $E_{N_{i}}^{*}$, $p_{N_{i}}^{*}$ are the energy and momentum of $N_{i}$ in rest frame of $M_{1}$ and $\gamma=\frac{E_{M_{1}}}{m_{M_{1}}}$.
$\theta_{N_{i}}^{*}$ is the emission
angle of particle $N_{i}$ in the rest frame of $M_{1}$, which is measured from the boost direction $\vec{\beta}$.
The energy  $E_{N_{i}}$ of the $N_{i}$ in the boosted $M_1$ frame lies within the range, 
\begin{align*}
{E_{N_{i}}}=\left[\big(\gamma E^*_{N_{i}}-{p_{N_{i}}^{*}}\sqrt{\gamma^{2}-1}\big),\big(\gamma E^*_{N_{i}}+{p_{N_{i}}^{*}}\sqrt{\gamma^{2}-1}\big)\right]
\end{align*}
Hence, the distribution of the energy of $N_{i}$ can be written as follows: 
\begin{align*}
 &\frac{1}{\Gamma\left(M_{1}^{+}\to\ell^{+}N_{i}\right)}\frac{d\Gamma\left(M_{1}^{+}\to\ell^{+}N_{i}\right)}{dE_{N_{i}}}=\frac{1}{2p_{N_{i}}^{*}\sqrt{\gamma^{2}-1}},
\end{align*}
The  signal event for $M^{+}_1 \to \ell^{+} \ell^{+} M^{-}_2$ in the lab-frame is: 
\begin{align}\label{event signal in flight decay}
N_{event}&\approx 2N_{M_{1}^{+}}\int_{E_{N_{i}}^{-}}^{E_{N_{i}}^{+}}dE_{N_{i}}\text{Br}\left(M_{1}^{+}\to\ell^{+}N_{i}\right)\nonumber\\
&\frac{m_{M_{1}}}{2p^{*}_{N_{i}}\left|\vec{p}_{M_{1}}\right|}\frac{\Gamma(N_{i}\to\ell^{+}M_{2}^{-})}{\Gamma_{N_{i}}}\mathcal{P}^{\prime}_{N},
\end{align}
where $\mathcal{P}^{\prime}_{N}=\left[1-exp\bigg(-\frac{M_{N_{i}}\Gamma_{N_{i}}L_{D}}{\sqrt{E_{N_{i}}^{2}-M_{N_{i}}^{2}}}\bigg)\right]$, is the probability of
$N_{i}$ to decay within the detector length $L_D$, after taking into account the boost factor. Since the LNV decays will be rare, the expected number of events for
these processes can be assumed to follow a Poisson distribution. Using the method of Feldman and Cousins, we get the average upper limit on the number of events at  
95$\%$ C.L., assuming zero background events and no true signal event to be $N_{event}=3.09$~\cite{Feldman}.

Note that the theoretical estimates of the number of events (given in Eqs.~(\ref{event signal in rest}) or~(\ref{event signal in flight decay}), corresponding to decay of parent meson at rest or in flight respectively) are functions of the mass parameters $M_{N_{i}}$ and $M_{W_{R}}$. If Majorana neutrinos having a mass such that they can be produced as an on shell intermediate state in the LNV meson decay processes exist, then, equating the numerical upper limit on the number of events to the theoretical expressions, results in constraints on $M_{W_{R}}$, corresponding to specific $M_{N_{i}}$ values for each of the
	following experiments.

\begin{itemize}
\item

\subsection{NA62}
NA62  is an ongoing experiment at CERN that will produce a large number of $K^{+}$ mesons~\cite{NA62 page}. 
The primary SPS 400 GeV proton beam, aims on a target, producing a secondary high intensity hadron beam with an optimum content of $K^{+}(\approx 6\%)$.
The expected number of $K^{+}$ decays in the fiducial volume is $4.5\times10^{12}$ per year.
Assuming three years of running, $N_{K^{+}}=1.35\times 10^{13}$. The detector length $L_{D}\approx 170$~m and the  produced $K^{+}$ mesons will decay in flight,
carrying a momentum of $75$ GeV. Non-observation of signal events for the decay mode $K^{+} \to \ell^{+} \ell^{+} \pi^{-}$ at NA62 can be used to set
limits on $M_{W_R}$ for $M_{N_{i}}~\sim 350$GeV. Using Eq.~(\ref{event signal in flight decay}), we derive the 95$\%$ C.L limit on $M_{W_R}$ for different
$M_{N_{i}}$ values, shown in Fig.~\ref{bounds on MWR for ee channel}(a) for the case of decay to like sign di-electrons.
From this ee channel, for a heavy Majorana neutrino mass $M_{N_{1}}\simeq 0.38$~GeV, the RH gauge boson mass can be constrained to be $M_{W_{R}}> 4.6$~TeV. For the $\mu\mu$ channel, 
for heavy neutrino mass $M_{N_{2}}\simeq 0.35$~GeV, a limit of $M_{W_{R}}> 4.3$~TeV can be obtained and is shown in Fig.~\ref{bounds on MWR for mu channel}(a).

\begin{figure}[]
\subfigure[]{\includegraphics[width=8cm]{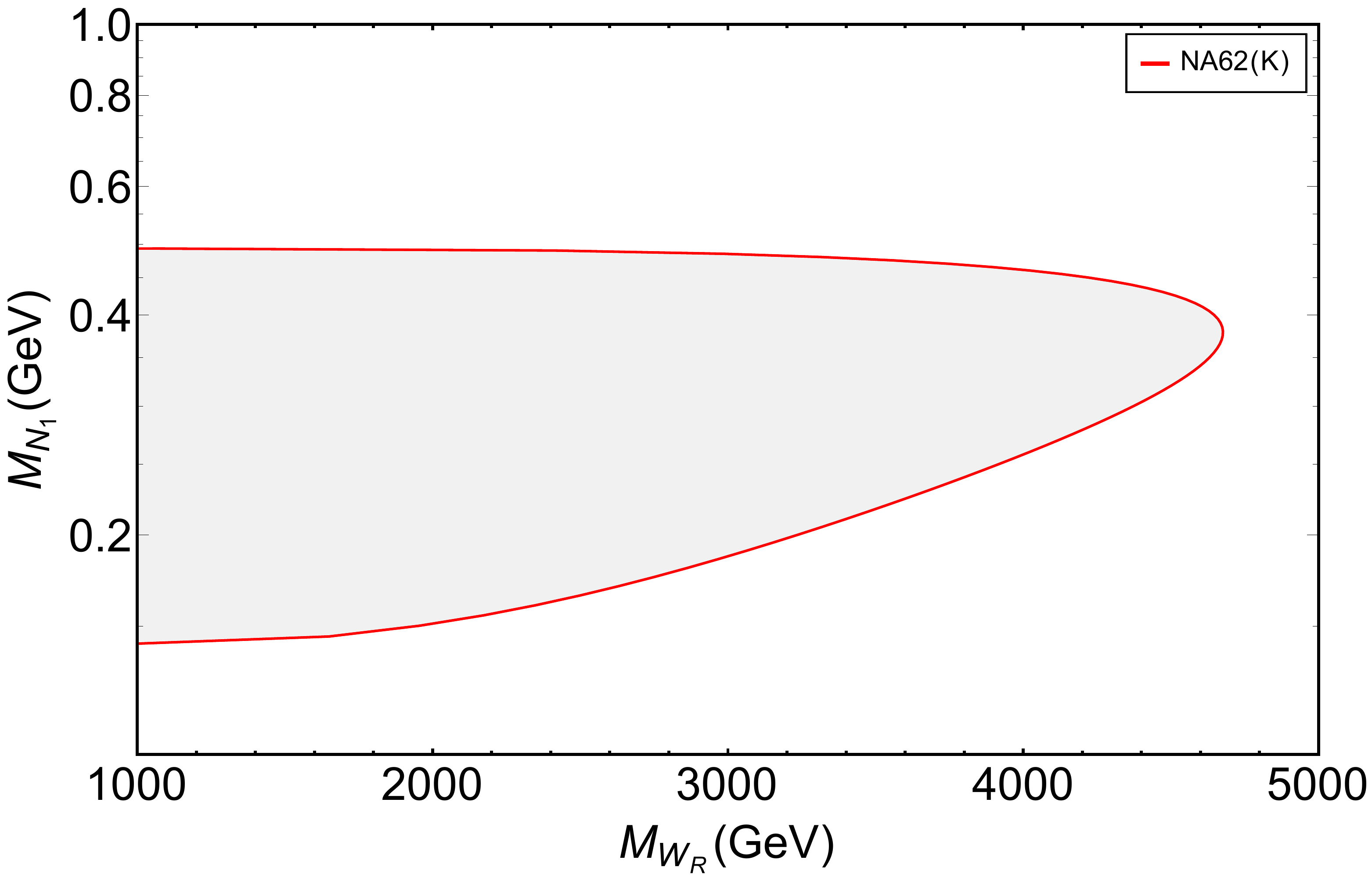}}
\subfigure[]{\includegraphics[width=8cm]{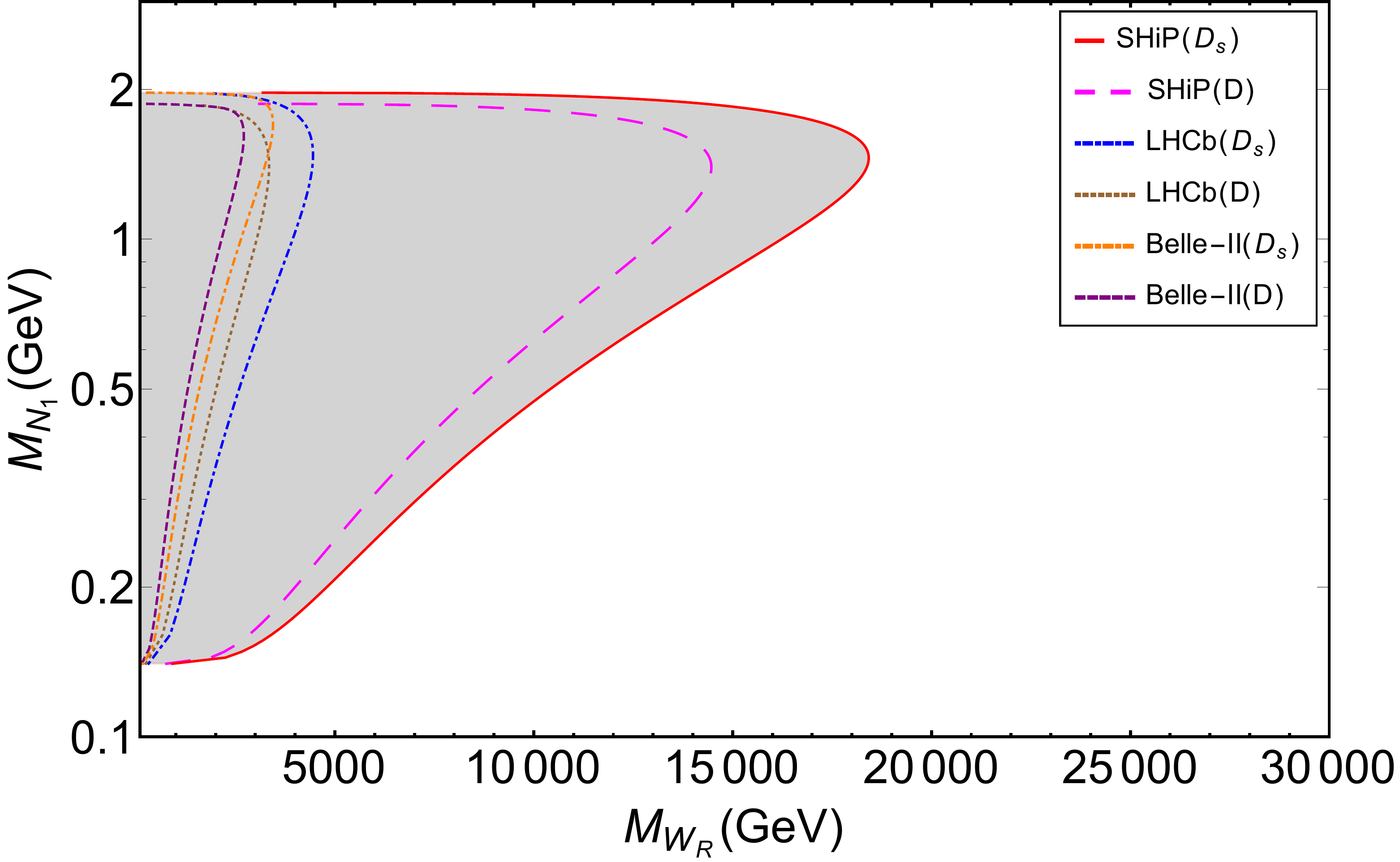}}
\subfigure[]{\includegraphics[width=8cm]{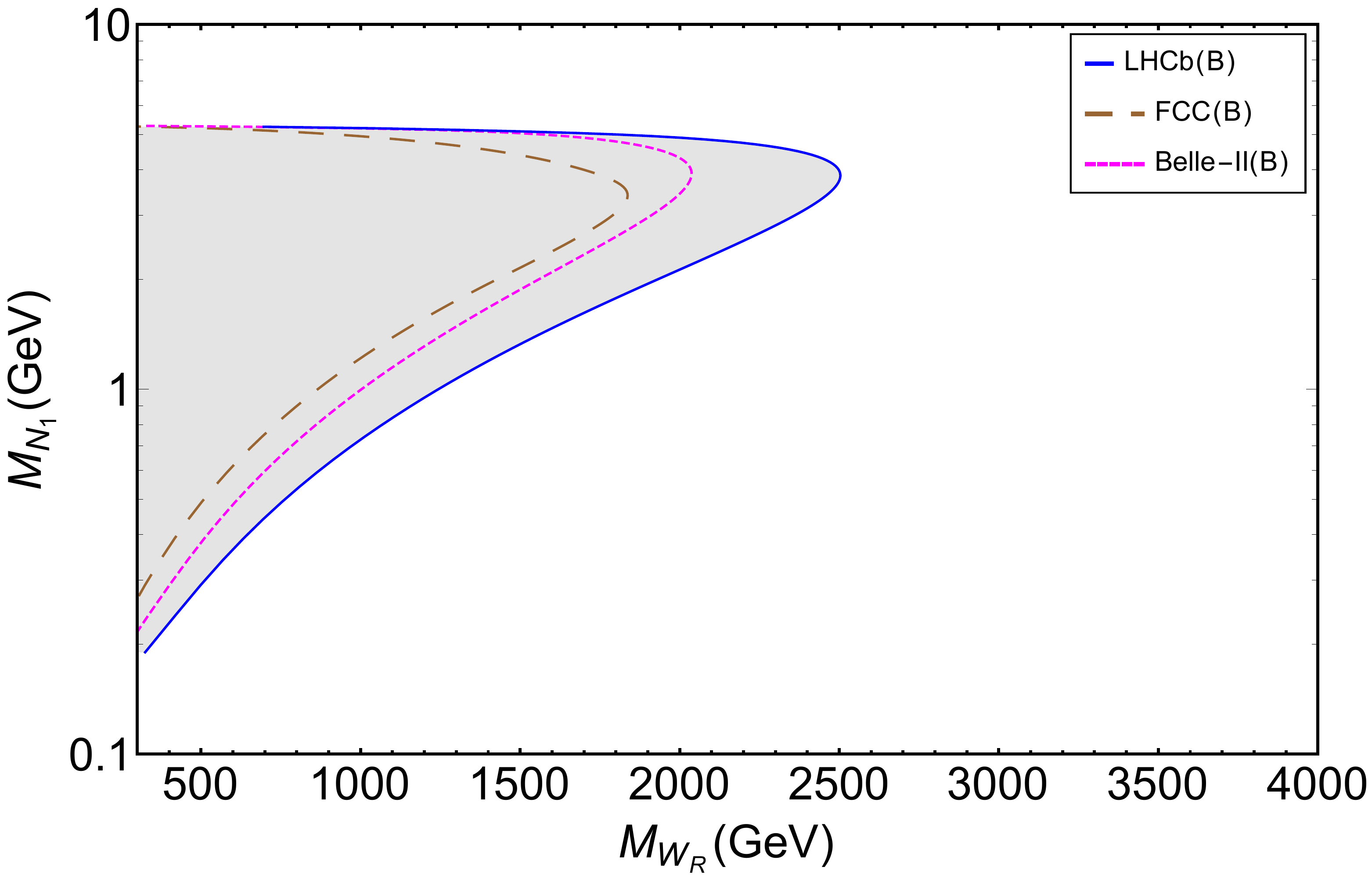}}
\caption{\small{Constraints on the RH gauge boson  $M_{W_{R}}$ mass, corresponding to a heavy neutrino $M_N$, that could be resonantly produced in lepton number
violating decays of (a)$K$, (b)$D,D_s$ and (c)$B$  mesons. The shaded region to the left of the curves corresponding to expected limits from searches at different ongoing
and future experiments will be ruled out in the absence of LNV meson decays with like sign di-electrons.}}
\label{bounds on MWR for ee channel}
\end{figure}

\item
\subsection{Belle II} 
The asymmetric SuperKEKB facility is designed to collide electron and positron beams such that the centre of mass energy is in the region of the $\Upsilon$ resonances.
An upgrade of Belle, the newly completed Belle II detector is  expected to collect data samples corresponding to an integrated
luminosity of 50 ab$^{-1}$ by the end of 2024~\cite{Belle 2}. The expected number of charged $B\bar{B}$ pairs to be produced at 50 ab$^{-1}$ is
{$5.5\times 10^{10}$}~\cite{PDG, Karim}. In addition, a large sample of charged $D, D_s$ mesons will also be accessible, with
$N_{D^+}=3.4\times 10^{10}$ and $N_{D_s^+}= 10^{10}$~\cite{Karim}.
A direct search for heavy Majorana neutrinos in B-meson decays was performed by the Belle collaboration using a data sample
that contained $772\times 10^{6}$
$B\bar{B}$ pairs~(at 711 fb$^{-1}$)~\cite{Belle search}.
At KEKB as well as superKEKB, the energies of the $e^{+}$, $e^{-}$ beam is sufficiently low so that the momentum of the produced $B$ mesons as well as that for the charmed mesons will not be appreciable and the suppression from high momentum
of the decaying mesons in the number of events will be absent.
Using the much larger expected number of mesons to be produced at Belle II ($N_{D^+},N_{D_s^+},N_{B^+}$) with the detector length  $L_{D}=1.5$~m, we calculate the expected
number of signal events for the LNV decays of
these mesons
using  Eq.~(\ref{event signal in rest}).
In Fig.~\ref{bounds on MWR for ee channel}(b),(c) and Fig. ~\ref{bounds on MWR for mu channel}(b),(c) we show the limits on $M_{W_R}$ and $M_{N_{i}}$
that will arise, if no events for the LNV $D^{+}, D^{+}_s, B^{+} \to \ell^{+}\ell^{+} \pi^{-}$ decays are observed. Note that, among the different decay modes, the most stringent
limit $M_{W_{R}}\geq 3.4$ TeV for heavy Majorana neutrino mass $M_{N_{1,2}}\simeq 1.7$ GeV can come from $D_{s}$ meson decays, which are Cabibbo favoured modes.

\item
\subsection{LHCb}
The LHCb detector is a forward spectometer at the Large Hadron Collider~(LHC) at CERN. During run 1, the LHCb detector collected data at $\sqrt{s}=7$~TeV with
integrated luminosity of $3~\text{fb}^{-1}$. During run 2, LHCb will collect additional $5~\text{fb}^{-1}$ at $\sqrt{s}=13$~TeV.
A search for heavy Majorana neutrinos
in B meson decays had been performed by the LHCb collaboration using the 7~TeV data~\cite{LHCb majorana search}.
The cross-section for producing B, $D$ and $D_{s}$ mesons at $\sqrt{s}=13$~TeV within the LHCb
acceptance~($2<\eta<5$) are 154 $\mu$b, 1000 $\mu$b and 460 $\mu$b respectively~\cite{B meson production, D and Ds meson production}. Hence, in run 2 with
$5~\text{fb}^{-1}$, expected number of B, $D$ and $D_{s}$ mesons are, $N_{B^{+}}=7.7\times 10^{11}$, $N_{D^{+}}=5\times 10^{12}$ and
$N_{D_{s}^{+}}=2.3\times 10^{12}$. The produced B and D mesons will decay in flight, carrying a momentum of order
of 100~GeV in forward direction~\cite{B momentum in LHCb}. We take the detector length $L_{D}\approx 20$~m.
The tightest constraint expected from LHCb are also from $D_{s}$ decays, $M_{W_{R}}\geq 4.4$ TeV for heavy Majorana neutrino mass $M_{N_1}\simeq 1.5$ GeV,
while for a heavier neutrino, $M_{N_1}\simeq 3.8$ GeV, the constraint is weaker, $M_{W_{R}}\geq 2.5$ TeV.

\item
\subsection{FCC-ee}
The Future Circular Collider (FCC-ee)~\cite{FCC} will  collect multi-ab$^{-1}$ integrated luminosities for $e^{+}e^{-}$ collisions at c.m.energy $\sqrt{s}\approx 91$~GeV. The expected number of $Z$-bosons is $10^{12}-10^{13}$. The number of charged B mesons from $Z$ decays can be estimated as,
\begin{align*}
N_{B^{+}}=N_{Z}\times \text{Br}\left(Z\to b\bar{b}\right)\times f_{u},
\end{align*}
where $N_{Z} \sim 10^{13}$, Br$\left(Z\to b\bar{b}\right)=0.1512$~\cite{pdg2014}, $f_{u}=0.410$~\cite{Heavy flavor group} is the fraction of $B^{+}$ from $\bar{b}$ quark
in $Z$ decay.
The B mesons produced at FCC-ee will have an energy distribution peaked at $E_{B^{+}}=\frac{M_{Z}}{2}$. Hence we can calculate the number
of signal events using Eq.~(\ref{event signal in flight decay}), where the detector length is taken to be, $L_{D}=2$~m. 
At FCC, for a heavy Majorana neutrino of mass $M_{N_{1,2}}\simeq 3.9$~GeV, RH gauge boson mass upto $M_{W_{R}}\simeq 2$~TeV
can be excluded using the decay modes $B^{+} \to \ell^{+}\ell^{+} \pi^{-}$.

\item
\subsection{SHiP}
The SHiP experiment is a newly proposed general purpose fixed target facility at the CERN SPS accelerator~\cite{ship}. A 400 GeV proton beam will be dumped on a heavy
target in order to produce $2\times 10^{20}$ proton-target interactions in five years. One of the goals of the experiment is to use decays of charmed mesons to search for
heavy sterile neutrinos using the decay mode $D^{+}_s/D^{+} \to \ell^{+} \ell^{+} \pi^{-}$. The number of charmed meson pairs that are expected to be produced in this experiment can be estimated as~\cite{SHiP physics paper},
\begin{align*}
 N_{meson}=X_{c\bar{c}}\times N_{POT}\times {\mathcal{R}},
\end{align*}
where $X_{c\bar{c}}$ is the $c\bar{c}$ production rate, $N_{POT}=2\times 10^{20}$ is the number of proton-target interaction. The relative abundances $\mathcal{R}$ of charmed mesons, such as, $D$ and $D_{s}$ are $30\%$ and $8\%$ respectively. Hence, the expected number of $D$ and $D_{s}$ mesons are $N_{D^{+}}=1.02\times 10^{17}$ and
$N_{D_{s}^{+}}=2.72\times10^{16}$ respectively. This very high intensity of the charmed mesons will permit the absence of LNV $D_s$ meson decay mode at SHiP,
to set tight constraints on the mass of RH gauge boson
$M_{W_{R}}> 18.4$~TeV ($M_{W_{R}}\sim 17.4$~TeV) for heavy Majorana neutrino mass $M_{N_{1}}\simeq 1.46$~GeV ($M_{N_{2}}\simeq 1.43$~GeV).
The detector length is taken to be, $L_{D}=60$~m. For the 400~GeV CNGS proton beam on target, the expected momentum of the produced charmed mesons is
$\sim$ 58~GeV~\cite{Gorbunov}.
The result for the constraints on $M_{W_{R}}$ are shown in Fig.~\ref{bounds on MWR for ee channel}(b) and Fig. ~\ref{bounds on MWR for mu channel}(b). 
\end{itemize}

\begin{figure}[]
\subfigure[]{\includegraphics[width=8cm]{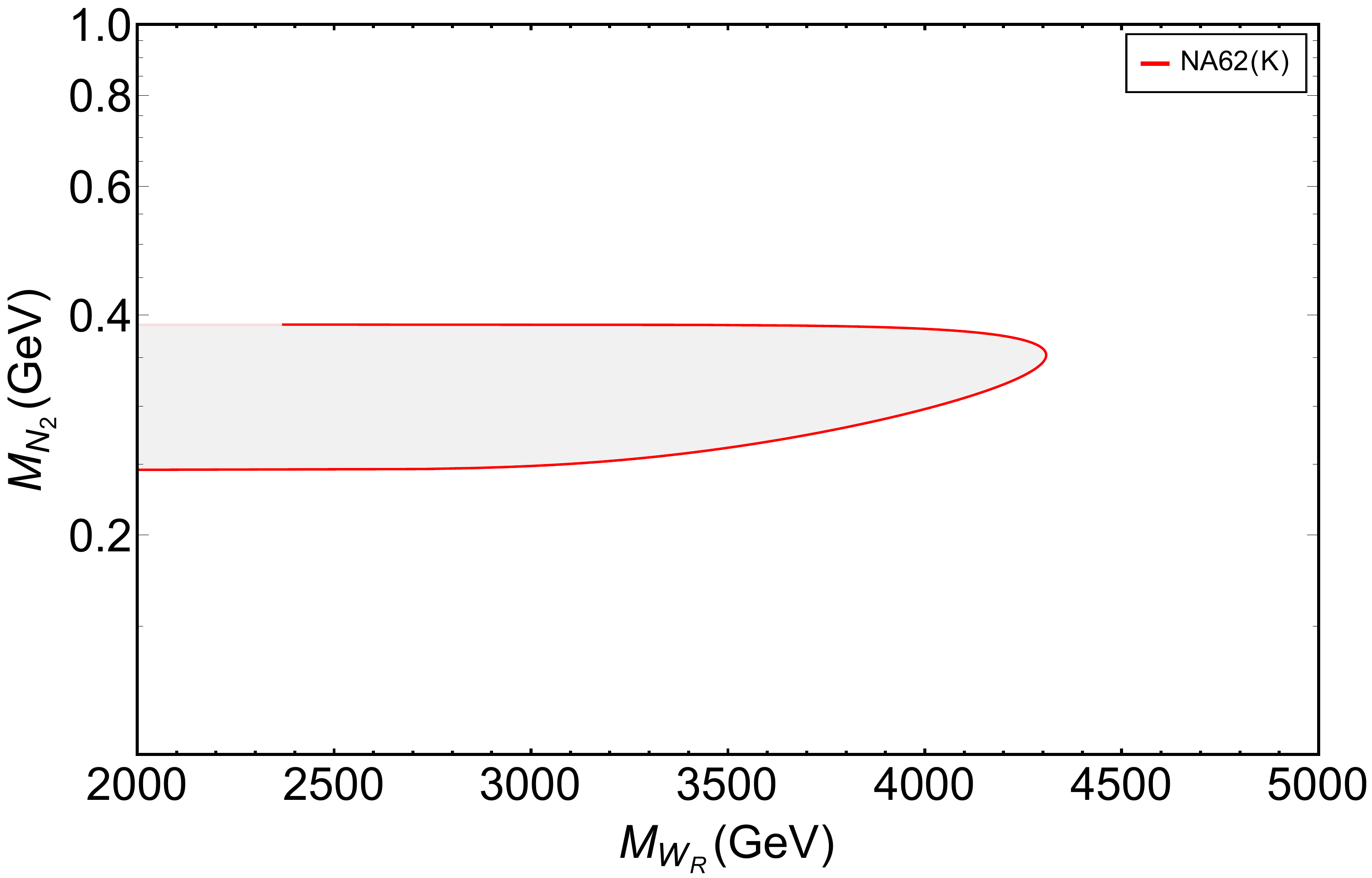}}
\subfigure[]{\includegraphics[width=8cm]{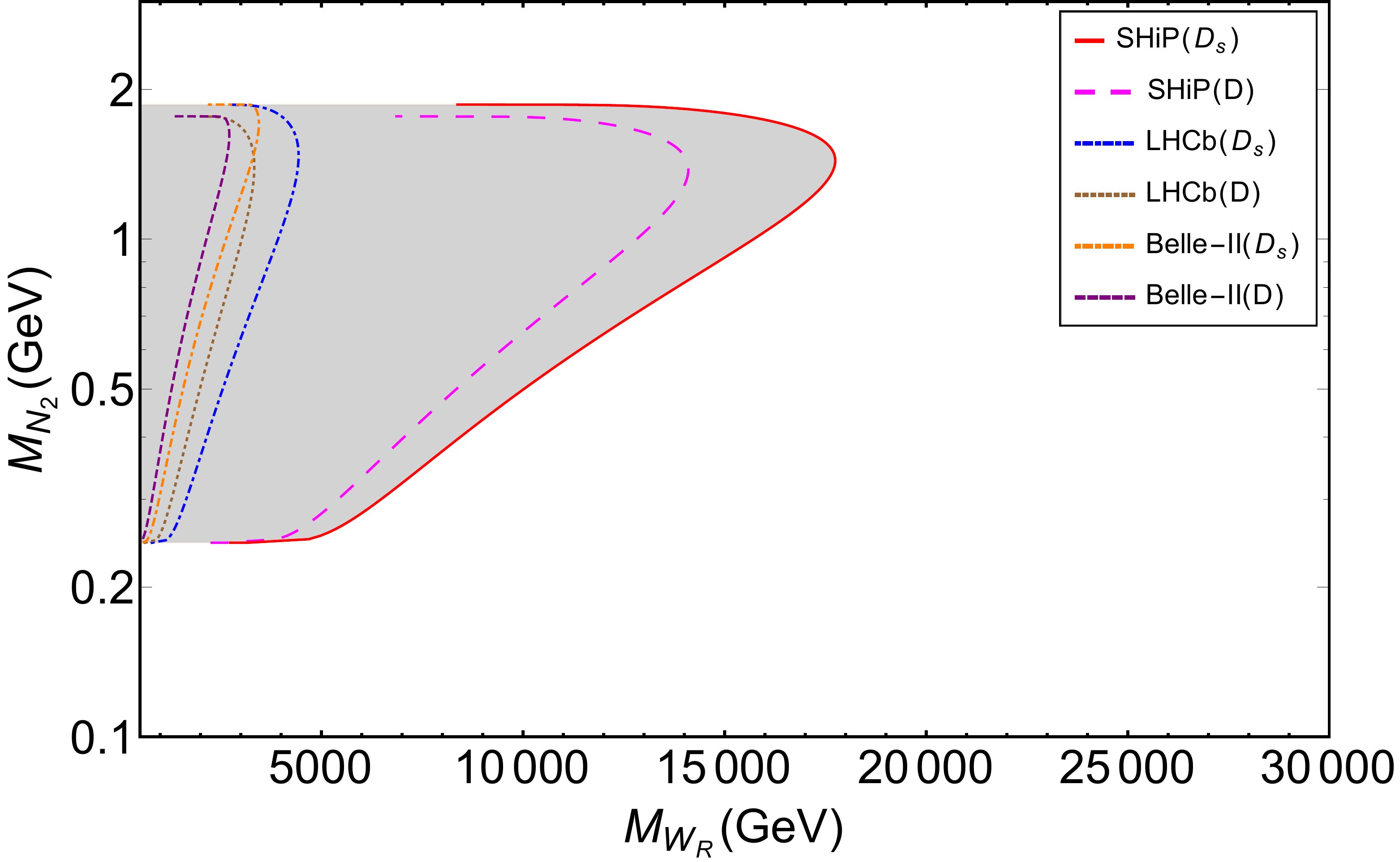}}
\subfigure[]{\includegraphics[width=8cm]{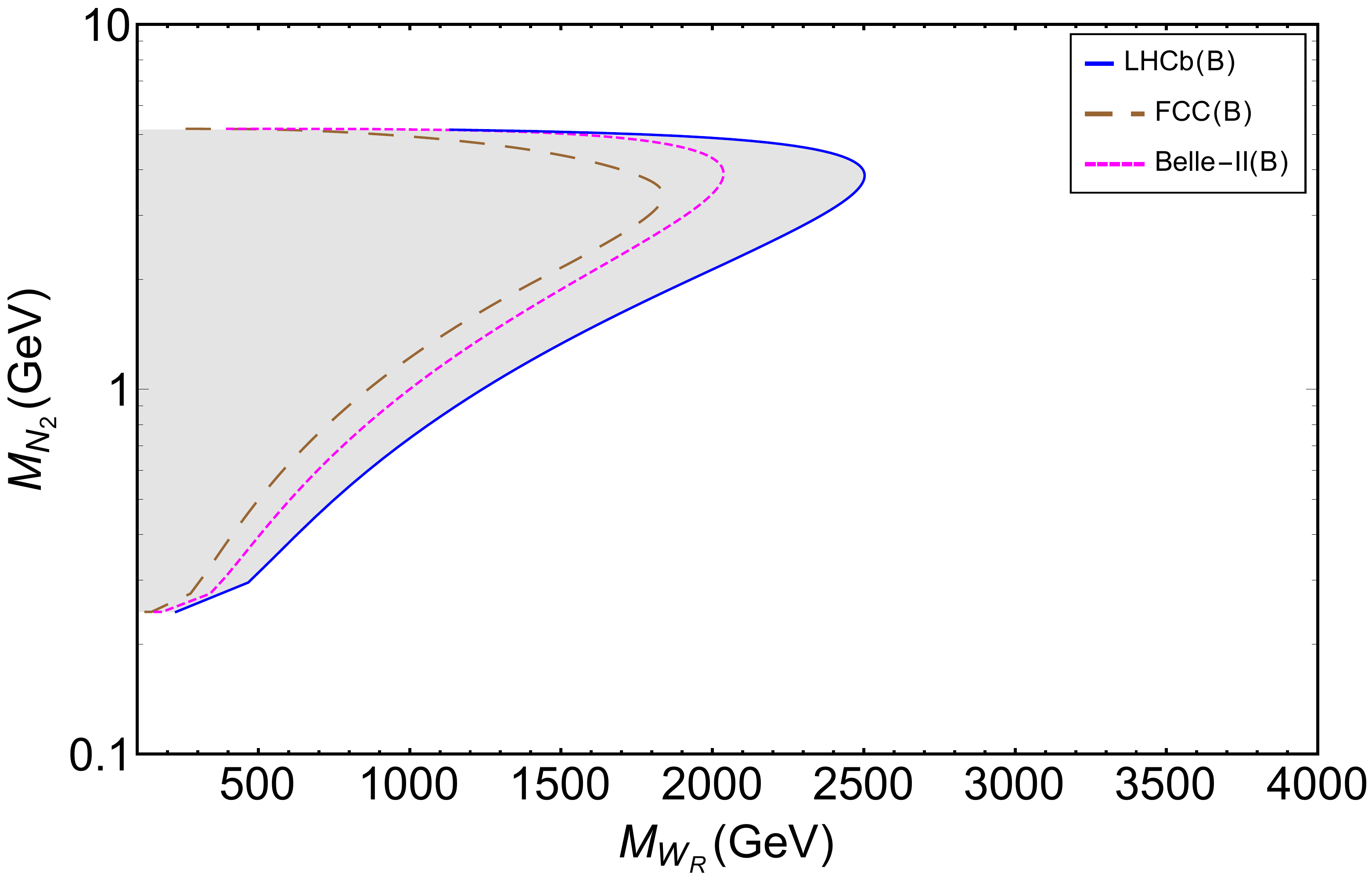}}
\caption{\small{Same as Fig.~\ref{bounds on MWR for ee channel}, except that the limits corresponding to searches at different ongoing and future experiments with
like sign dimuons are shown.}}
\label{bounds on MWR for mu channel}
\end{figure}

From Fig.~\ref{bounds on MWR for ee channel} and Fig.~\ref{bounds on MWR for mu channel} it is evident that the most stringent limit in the $M_N \sim 1$ GeV range will be provided by
SHiP with the $D_s/D \to \ell^{+} \ell^{+} \pi^{-}$ decay mode. In the
relatively higher mass range $M_N \sim 4$ GeV, stringent limit can be obtained by FCC-ee, Belle-II and LHCb experiments. Besides,
the ongoing NA62 can give constraint $M_{W_R}>4.6$~TeV($M_{N_{1}}\sim 0.38$~GeV), which is competitive with the collider bounds from LHC.
Note that our numerical constraints on $M_{W_R}$ for various experiments are obtained assuming idealized detectors with $100\%$ detection,
reconstruction efficiencies etc. The realistic constraints (expected to be weaker) will only be feasible through searches by the experimental collaborations,
incorporating all these corrections.

\section{Comparison with Existing constraints from other experiments}

In addition to the meson decay searches, there are other direct and indirect searches that constrain the mass of heavy neutrino $N$ and $M_{W_R}$.
In particular, direct collider searches, such as,  LHC  dijet searches  for $W^{\prime}$ \cite{ATLAS:2016lvi}, the same sign dilepton searches
\cite{Aad:2015xaa, Khachatryan:2014dka} and indirect searches, such as, $0\nu \beta \beta$ and LFV $\mu \to e \gamma$, 
give stringent constraints on the masses of the   gauge boson $W_R$ and heavy neutrino $N$. Assuming a 75$\%$ branching
ratio of $W_R \to j j$, the  13 TeV ATLAS
dijet search ruled out the $W_R$ mass upto 2.9 TeV  \cite{ATLAS:2016lvi}. For $W_R$ that couples to two light generation
of quarks through CKM type mixing, the branching
ratios to of $W_R \to j j $ is  $60 \%$ \cite{Mitra:2016kov}. The limit will be comparable with the reported limit
from ATLAS.  On the other hand, the search for
same-sign dilepton at LHC is only relevant for heavy neutrino mass $M_N$ in the 100 GeV-TeV range. The $95 \%$ C.L limit
from ATLAS 8 TeV search on the $M_{W_R}$
reaches 2.9 TeV \cite{Aad:2015xaa, Helo:2015ffa}.  For all the mediators $M_{W_R}$ and $M_N$ in the TeV-few hundred GeV
mass range, the LHC same sign lepton search is more constraining than $0 \nu \beta \beta$ \cite{Deppisch:2012nb}.
The latter is sensitive to a wide  range of heavy neutrino $N$ and RH gauge boson $W_R$ masses. The tightest bound
from $0\nu \beta \beta$ $M_{W_R} > 9-10$ TeV is applicable for $M_N \sim 0.1$ GeV \cite{Helo:2015ffa}. The meson decays, on the other hand, are sensitive 
in constraining the low mass region of RH neutrino of mass between few hundred MeV to few GeV. The ongoing experiment NA62 can
constrain $M_{W_R}>4.6$~TeV($M_{N_{1}}\sim 0.38$~GeV), that will be more stringent than the limit provided by LHC-dijet search. The other future  experiments,
such as, SHiP can allow upto a very large mass $M_{W_R}> 18.4$~TeV($M_{N_{1}}\sim 1.46$~GeV).

\section{conclusions}

We evaluate the  lepton number violating meson decays $M_1 \to \ell^{+} \ell^{+} M_2$ within the framework of a Left-Right symmetric model. The right handed Majorana neutrinos of masses in the $\sim (100\, \rm{MeV}-5\, \rm{GeV})$ range, can result in a resonant enhancement of these processes. These neutrinos along with the left handed and right handed gauge bosons mediate these processes, with contributions from  $W_R-N_{i}-W_R$, $W_L-N_{i}-W_L$, $W_R-N_{i}-W_L$ and $W_L-N_{i}-W_R$ exchanges. If Majorana neutrinos in this low mass range ($\sim$ upto few GeV), exist, then, non-observation of the LNV meson decays at the various ongoing and future experiments will result in constraints
	on the RH gauge boson $M_{W_{R}}$, corresponding to the Majorana neutrino mass $M_{N_{i}}$.
The ongoing experiment NA62 can provide
the limit $M_{W_R}>4.6$~TeV($M_{N_{1}}\sim 0.38$~GeV), that is more
stringent than the present collider constraint on $W_R$. The future experiment, such as SHiP will be sensitive upto a very large mass $M_{W_R}> 18.4$~TeV($M_{N_{1}}\sim 1.46$~GeV) which will be tighter than any collider constraint but will correspond to a low value of $M_N$. The meson decays are sensitive for  low mass right handed neutrinos (in the few $100$ MeV-few GeV range) and are  complementary to LHC (sensitive to few hundred GeV to TeV mass neutrinos).

\section{Appendix}
The different partial decay widths of the RH neutrinos $N_{i}$ are 
\begin{widetext}
\begin{align*}
\Gamma (N_{j}\rightarrow\ell^{-}P^{+})&=\frac{G_{F}^{2}M_{N_{j}}^{3}}{16\pi}f_{p}^{2}\left|V_{q\bar{q}^{'}}\right|^{2}\bigg(\left|S_{\ell_1 N_{j}}\right|^{2}F_{P}\left(x_{\ell},x_{P}\right)+\left|V_{\ell_1 N_{j}}\right|^{2}\xi_{1}^{4}F_{P}
 \left(x_{\ell},x_{P}\right)\\
& +4Re\left[S_{\ell_1N_{j}}V_{\ell_1N_{j}}\right]\xi_{1}^{2}x_{\ell}x_{P}^{2}\lambda^{\frac{1}{2}}\left(1,x_{\ell}^{2},x_{P}^{2}\right)\bigg);\\
\Gamma\left(N_{j}\rightarrow\ell^{-}V^{+}\right)&=\frac{G_{F}^{2}M_{N_{j}}^{3}}{16\pi}f_{V}^{2}\left|V_{q\bar{q}^{'}}\right|^{2}\bigg(\left|S_{\ell_1N_{j}}\right|^{2}F_{V}\left(x_{\ell},x_{V}\right)+
 \left|V_{\ell_1N_{j}}\right|^{2}\xi_{1}^{4}F_{V}\left(x_{\ell},x_{V}\right)\\
&-12Re\left[S_{\ell_1N_{j}}V_{\ell_1N_{j}}\right]\xi_{1}^{2}x_{\ell}x_{V}^{2}\lambda^{\frac{1}{2}}\left(1,x_{\ell}^{2},x_{V}^{2}\right)\bigg);\\
 \Gamma\left(N_{j}\rightarrow\nu_{\ell}P^{0}\right)&=\frac{G_{F}^{2}M_{N_{j}}^{3}}{4\pi}f_{P}^{2}\sum_{i}\left|U_{\ell i}\right|^{2}\left|S_{\ell_1N_{j}}\right|^{2}\bigg(K_{P}^{2}
 +K_{P}^{'2}\xi_{2}^{4}-2K_{P}K_{P}^{'}\xi_{2}^{2}\bigg)F_{P}\left(x_{\nu_{\ell}},x_{P}\right);\\
\Gamma\left(N_{j}\rightarrow\nu_{\ell}V^{0}\right)&=\frac{G_{F}^{2}M_{N_{j}}^{3}}{4\pi}f_{V}^{2}\sum_{i}\left|U_{\ell i}\right|^{2}\left|S_{\ell_1N_{j}}\right|^{2}\bigg(K_{V}^{2}
 +K_{V}^{'2}\xi_{2}^{4}-2K_{V}K_{V}^{'}\xi_{2}^{2}\bigg)F_{V}\left(x_{\nu_{\ell}},x_{P}\right);\\
 \end{align*}
 \begin{align*}
 \Gamma\left(N_{j}\rightarrow\ell_{1}^{-}\ell_{2}^{+}\nu_{\ell_{2}}\right)&=\frac{G_{F}^{2}M_{N_{j}}^{5}}{16\pi^{3}}\bigg(\left|S_{\ell_{1}N_{j}}\right|^{2}\sum_{i}\left|U_{\ell_{2}i}\right|^{2}
 I_{1}\left(x_{\ell_{1}},x_{\nu_{\ell_{2}}},x_{\ell_{2}}\right)+\left|V_{\ell_{1}N_{j}}\right|^{2}\sum_{i}\left|T_{\ell_{2}i}\right|^{2}\xi_{1}^{4}
 I_{1}\left(x_{\ell_{1}},x_{\nu_{\ell_{2}}},x_{\ell_{2}}\right)\\
 &-8Re\big(S_{\ell_{1}N_{j}}^{*}V_{\ell_{1}N_{j}}^{*}\sum_{i}U_{\ell_{2}i}T_{\ell_{2}i}\big)\xi_{1}^{2}I_{3}\left(x_{\ell_{1}},x_{\nu_{\ell_{2}}},x_{\ell_{2}}\right)\bigg);\\
\Gamma\left(N_{j}\rightarrow\nu_{\ell_{2}}\ell_{2}^{-}\ell_{2}^{+}\right)&=\frac{G_{F}^{2}M_{N_{j}}^{5}}{16\pi^{3}}\bigg(\left|S_{\ell_{2}N_{j}}\right|^{2}\sum_{i}\left|U_{\ell_{2}i}\right|^{2}\bigg[I_{1}
 \left(x_{\nu_{\ell_{2}}},x_{\ell_{2}},x_{\ell_{2}}\right)+2\left((g_{V}^{\ell})^{2}+(g_{A}^{\ell})^{2}\right)I_{1}
 \left(x_{\nu_{\ell_{2}}},x_{\ell_{2}},x_{\ell_{2}}\right)\\
 &+2\left((g_{V}^{\ell})^{2}-(g_{A}^{\ell})^{2}\right)I_{2}
 \left(x_{\nu_{\ell_{2}}},x_{\ell_{2}},x_{\ell_{2}}\right)+2\left((g_{V}^{\prime\ell})^{2}+(g_{A}^{\prime\ell})^{2}\right)\xi_{2}^{4}I_{1}
 \left(x_{\nu_{\ell_{2}}},x_{\ell_{2}},x_{\ell_{2}}\right)\nonumber\\
 &+2\left((g_{V}^{\prime\ell})^{2}-(g_{A}^{\prime\ell})^{2}\right)\xi_{2}^{4}I_{2}
 \left(x_{\nu_{\ell_{2}}},x_{\ell_{2}},x_{\ell_{2}}\right)
 -4\xi_{2}^{2}\big((g_{V}^{\ell}g_{V}^{\prime\ell}+g_{A}^{\ell}g_{A}^{\prime\ell})I_{1}
 \left(x_{\nu_{\ell_{2}}},x_{\ell_{2}},x_{\ell_{2}}\right)\\
 &+(g_{V}^{\ell}g_{V}^{\prime\ell}-g_{A}^{\ell}g_{A}^{\prime\ell})I_{2}
 \left(x_{\nu_{\ell_{2}}},x_{\ell_{2}},x_{\ell_{2}}\right)\big)\bigg]
+\left|V_{\ell_{2}N_{j}}\right|^{2}\sum_{i}\left|T_{\ell_{2}i}\right|^{2}\xi_{1}^{4}I_{1}
 \left(x_{\nu_{\ell_{2}}},x_{\ell_{2}},x_{\ell_{2}}\right)\\
 &-8Re\big[S_{\ell_{2}N_{j}}^{*}V_{\ell_{2}N_{j}}^{*}\sum_{i}U_{\ell_{2}i}T_{\ell_{2}i}\big]\xi_{1}^{2}
 I_{3}\left(x_{\ell_{2}},x_{\nu_{\ell_{2}}},x_{\ell_{2}}\right)+2Re\big[\left|S_{\ell_{2}N_{j}}\right|^{2}\sum_{i}\left|U_{\ell_{2}i}\right|^{2}\big]
 \bigg[\xi_{2}^{2}(g_{A}^{\prime\ell}-g_{V}^{\prime\ell})I_{1}
 \left(x_{\nu_{\ell_{2}}},x_{\ell_{2}},x_{\ell_{2}}\right)\nonumber\\
& -\xi_{2}^{2}(g_{A}^{\prime\ell}+g_{V}^{\prime\ell})I_{2}
 \left(x_{\nu_{\ell_{2}}},x_{\ell_{2}},x_{\ell_{2}}\right)-(g_{A}^{\ell}-g_{V}^{\ell})I_{1}
 \left(x_{\nu_{\ell_{2}}},x_{\ell_{2}},x_{\ell_{2}}\right)
 +(g_{A}^{\ell}+g_{V}^{\ell})I_{2}
 \left(x_{\nu_{\ell_{2}}},x_{\ell_{2}},x_{\ell_{2}}\right)\bigg]\nonumber\\
&-8Re\big[S_{\ell_{2}N_{j}}V_{\ell_{2}N_{j}}\sum_{i}U_{\ell_{2}i}^{*}T_{\ell_{2}i}^{*}\big]\xi_{1}^{2}\bigg[(g_{V}^{\prime\ell}-g_{A}^{\prime\ell})\xi_{2}^{2}I_{3}\left(x_{\nu_{\ell_{2}}},x_{\ell_{2}},x_{\ell_{2}}\right)
+\frac{1}{4}(g_{V}^{\prime\ell}+g_{A}^{\prime\ell})\xi_{2}^{2}I_{4}\left(x_{\ell_{2}},x_{\ell_{2}},x_{\nu_{\ell_{2}}}\right)\nonumber\\
&+(g_{V}^{\ell}-g_{A}^{\ell})I_{3}\left(x_{\nu_{\ell_{2}}},x_{\ell_{2}},x_{\ell_{2}}\right)
+\frac{1}{4}(g_{V}^{\ell}+g_{A}^{\ell})I_{4}\left(x_{\ell_{2}},x_{\ell_{2}},x_{\nu_{\ell_{2}}}\right)\bigg]\bigg);\\
\Gamma\left(N_{j}\rightarrow\nu_{\ell_{1}}\ell_{2}^{-}\ell_{2}^{+}\right)&=\frac{G_{F}^{2}M_{N_{j}}^{5}}{8\pi^{3}}\left|S_{\ell_{1}N_{j}}\right|^{2}\sum_{i}\left|U_{\ell_{1}i}\right|^{2}\bigg[\big((g_{V}^{\ell})^{2}
+(g_{A}^{\ell})^{2}\big)I_{1} \left(x_{\nu_{\ell_{1}}},x_{\ell_{2}},x_{\ell_{2}}\right)
 +\big((g_{V}^{\ell})^{2}-(g_{A}^{\ell})^{2}\big)I_{2} \left(x_{\nu_{\ell_{1}}},x_{\ell_{2}},x_{\ell_{2}}\right)\\
&+\big((g_{V}^{\prime\ell})^{2}
+(g_{A}^{\prime\ell})^{2}\big)\xi_{2}^{4}I_{1} \left(x_{\nu_{\ell_{1}}},x_{\ell_{2}},x_{\ell_{2}}\right)
+\big((g_{V}^{\prime\ell})^{2}-(g_{A}^{\prime\ell})^{2}\big)\xi_{2}^{4}I_{2} \left(x_{\nu_{\ell_{1}}},x_{\ell_{2}},x_{\ell_{2}}\right)\\
&-2\xi_{2}^{2}\big[(g_{V}^{\ell}g_{V}^{\prime\ell}+g_{A}^{\ell}g_{A}^{\prime\ell})
I_{1} \left(x_{\nu_{\ell_{1}}},x_{\ell_{2}},x_{\ell_{2}}\right)+(g_{V}^{\ell}g_{V}^{\prime\ell}-g_{A}^{\ell}g_{A}^{\prime\ell})
I_{2}\left(x_{\nu_{\ell_{1}}},x_{\ell_{2}},x_{\ell_{2}}\right)\big]\bigg].
\end{align*}
In the above decay mode $\ell_{1}\ne\ell_{2}$.
\begin{align*}
\Gamma\left(N_{j}\rightarrow\nu_{\ell}\nu\overline{\nu}\right)&=\frac{G_{F}^{2}M_{N_{j}}^{5}}{192\pi^{3}}\left|S_{\ell N_{j}}\right|^{2}\sum_{i}\left|U_{\ell i}\right|^{2}
\bigg(1-\text{sin}^{2}\theta_{w}\xi_{2}^{2}\bigg)^{2},\\
\end{align*}
where $\xi_{1}=\frac{M_{W_{L}}}{M_{W_{R}}}$, $\xi_{2}=\frac{M_{Z}}{M_{Z'}}$, $x_{i}=\frac{m_{i}}{M_{N}}$ with $m_{i}=m_{\ell}, m_{P^{0}},m_{V^{0}},m_{P^{+}},m_{V}^{+}$.
The kinematical function are given by,
\begin{align*}
I_{1}(x,y,z)&=\int_{(x+y)^{2}}^{(1-z)^{2}}\frac{ds}{s}(s-x^{2}-y^{2})(1+z^{2}-s)\lambda^{\frac{1}{2}}(s,x^{2},y^{2})\lambda^{\frac{1}{2}}(1,s,z^{2});\\
I_{2}(x,y,z)&=yz\int_{(y+z)^{2}}^{(1-x)^{2}}\frac{ds}{s}(1+x^{2}-s)\lambda^{\frac{1}{2}}(s,y^{2},z^{2})\lambda^{\frac{1}{2}}(1,s,x^{2});\\
I_{3}(x,y,z)&=xyz\int_{(x+y)^{2}}^{(1-z)^{2}}\frac{ds}{s}\lambda^{\frac{1}{2}}(s,x^{2},y^{2})\lambda^{\frac{1}{2}}(1,s,z^{2});\\
I_{4}(x,y,z)&=z\int_{(x+y)^{2}}^{(1-z)^{2}}\frac{ds}{s}\lambda^{\frac{1}{2}}(s,x^{2},y^{2})\lambda^{\frac{1}{2}}(1,s,z^{2});\\
F_{P}(x,y)&=\big((1+x^{2})(1+x^{2}-y^{2})-4x^{2}\big)\lambda^{\frac{1}{2}}(1,x^{2},y^{2});\\
F_{V}(x,y)&=\big((1-x^{2})^{2}+(1+x^{2})y^{2}-2y^{4}\big)\lambda^{\frac{1}{2}}(1,x^{2},y^{2}).
\end{align*}
Neutral current couplings of leptons are given by,
\begin{eqnarray}
 &g_{V}^{\ell}=-\frac{1}{4}+\text{sin}^{2}\theta_{w},\,\,\,g_{A}^{\ell}=\frac{1}{4},\nonumber\\
 &g_{V}^{\prime\ell}=-\frac{1}{4}+\text{sin}^{2}\theta_{w},\,\,\,g_{A}^{\prime\ell}=-\frac{1}{4}+\frac{1}{2}\text{sin}^{2}\theta_{w}.\nonumber
\end{eqnarray}
Neutral current coupling of pseudoscalar mesons are given by,
\begin{eqnarray}
 &K_{\pi^{0}}=-\frac{1}{2\sqrt{2}},\,\,\,K_{\pi^{0}}^{\prime}=\frac{1}{\sqrt{2}}(\frac{1}{2}-\text{sin}^{2}\theta_{w}),\nonumber\\
 &K_{\eta}=-\frac{1}{2\sqrt{6}},\,\,\,K_{\eta}^{\prime}=\frac{1}{\sqrt{6}}(\frac{1}{2}-\text{sin}^{2}\theta_{w}),\nonumber\\
 &K_{\eta^{\prime}}=\frac{1}{4\sqrt{3}},\,\,\,K_{\eta^{\prime}}^{\prime}=\frac{1}{\sqrt{3}}(-\frac{1}{4}+\frac{1}{2}\text{sin}^{2}\theta_{w}),\nonumber\\
 &K_{\eta_{c}}=-\frac{1}{4},\,\,\,K_{\eta_{c}}^{\prime}=(\frac{1}{4}-\frac{1}{2}\text{sin}^{2}\theta_{w}),\nonumber
\end{eqnarray}
Neutral current coupling of vector mesons are given by,
\begin{eqnarray}
 &K_{\rho^{0}}=\frac{1}{\sqrt{2}}(\frac{1}{2}-\text{sin}^{2}\theta_{w}),\nonumber\\
 &K_{\omega}=-\frac{1}{3\sqrt{2}}\text{sin}^{2}\theta_{w},\nonumber\\
 &K_{\phi}=(-\frac{1}{4}+\frac{1}{3}\text{sin}^{2}\theta_{w}),\nonumber\\
 &K_{J/\psi}=(\frac{1}{4}-\frac{2}{3}\text{sin}^{2}\theta_{w}).\nonumber
\end{eqnarray}
\end{widetext}

\section*{Acknowledgments}
MM acknowledges the supports from Institute of
Physics (IOP) and DST INSPIRE Research Grant
(INSPIRE-15-0074).


\end{document}